\documentclass[journal]{IEEEtran}
\usepackage{indentfirst}
\usepackage{threeparttable}
\usepackage{booktabs}
\usepackage{setspace}
\usepackage{graphicx}
\usepackage{subfigure}
\usepackage{booktabs}
\usepackage{array}
\usepackage{cite}
\usepackage{makecell}
\usepackage{algorithm,algorithmic}
\usepackage{caption}
\usepackage{epsfig}
\usepackage{xcolor}

\usepackage{multirow}
\usepackage{amsfonts}
\usepackage{caption}

\newcommand{\bm}[1]{\mbox{\boldmath{$#1$}}}

\ifCLASSINFOpdf
\else

\fi

\usepackage{algorithmic}

\begin{document}
%
\title{Structured Distributed Compressive Channel Estimation over Doubly Selective Channels}

\author{\IEEEauthorblockN{Qibo~Qin, Lin~Gui, \emph{Member,~IEEE}, Bo~Gong, Xiang~Ren, and Wen~Chen, \emph{Senior Member,~IEEE}}


\thanks{This work is supported in part by the National Natural Science Foundation of China 61471236, and by National 973 project under grant 2012CB316106.

Q. Qin, L. Gui, B. Gong and X. Ren are with the Department of Electronic Engineering, Shanghai Jiao Tong University, Shanghai 200240, China (e-mail:qinqibo@sjtu.edu.cn).

W. Chen is with Shanghai Key Laboratory of Navigation and Location based Services, Shanghai Jiao Tong University, and the School of Electronics Engineering and Automation, Guilin University of Electronics Technology.
}}

\maketitle

\begin{abstract}
For an orthogonal frequency-division multiplexing (OFDM) system over a doubly selective (DS) channel, a large number of pilot subcarriers are needed to estimate the numerous channel parameters, resulting in low spectral efficiency. In this paper, by exploiting temporal correlation of practical wireless channels, we propose a highly efficient structured distributed compressive sensing (SDCS) based joint multi-symbol channel estimation scheme. Specifically, by using the complex exponential basis expansion model (CE-BEM) and exploiting the sparsity in the delay domain within multiple OFDM symbols, we turn to estimate jointly sparse CE-BEM coefficient vectors rather than numerous channel taps. Then a sparse pilot pattern within multiple OFDM symbols is designed to obtain an ICI-free structure and transform the channel estimation problem into a joint-block-sparse model. Next, a novel block-based simultaneous orthogonal matching pursuit (BSOMP) algorithm is proposed to jointly recover coefficient vectors accurately. Finally, to reduce the CE-BEM modeling error, we carry out smoothing treatments of already estimated channel taps via piecewise linear approximation. Simulation results demonstrate that the proposed channel estimation scheme can achieve higher estimation accuracy than conventional schemes, although with a smaller number of pilot subcarriers.
\end{abstract}


\IEEEpeerreviewmaketitle

\section{Introduction}

\IEEEPARstart{O}{rthogonal} frequency division multiplexing (OFDM) is an attractive communication system because of its robustness against frequency-selective (FS) fading channels, high data rate transmission capability and high spectral efficiency. Recently, OFDM has gained its popularity in a number of wireless broadband communication systems, such as the digital video broadcasting (DVB) system, IEEE802.16e (WIMAX), and 3GPP long-term evolution (LTE) systems \cite{habibsenol,Jin2015,liangzhang2014}. As the accurate channel state information (CSI) can notably improve system performance, it is necessary to provide a reliable channel estimation method \cite{GuanGuiBayesian}. Most existing researches consider the FS channel with slow time-variation properties, and channel estimation has not posed a severe challenge \cite{pengchengTcom}.

Unfortunately, OFDM systems are sensitive to Doppler effect induced by fast time variations, which will destroy the orthogonality among subcarriers and induce inter-carrier interference (ICI). The channels faced with both FS and time-selective (TS) fading are often referred to as doubly-selective (DS) channels \cite{learningDS2008}. Channel estimation for DS channels is extremely challenging due to the fact that the parameters to be estimated are numerous. For example, the total number of unknown channel parameters within a single OFDM symbol is \emph{NL}, where \emph{N} is the number of subcarriers and \emph{L} is the maximum delay spread of the channel impulse responses (CIR) and both are very large in many broadband systems \cite{pengchengTcom}. To estimate the numerous channel parameters, lots of pilot subcarriers are required, resulting in a low spectral efficiency.

  The recently introduced methodology of compressive sensing (CS) is capable of reconstructing sparse signals from fewer samples than what is required by Nyquist rate \cite{xueyunheTVT}. Further, growing experimental studies verify that many wireless broadband channels exhibit sparsity, where the delay spread could be very large but the number of channel taps with significant power is usually small \cite{CCSwaheed}, \cite{GuanGuiHigh}. As such, applying CS theory to the OFDM channel estimation can dramatically reduce the number of pilot subcarriers. In \cite{Georg2008,xiangrenGlobe,Georgleakage2010, XiangRenTVT}, CS theory has been applied to estimate a DS channel, showing better estimation performance than conventional channel estimation methods such as minimal mean square error (MMSE) and least squares (LS) methods. However, in \cite{Georg2008}, \cite{xiangrenGlobe}, the ICI is treated as noise, and in \cite{Georgleakage2010}, \cite{XiangRenTVT}, an iterative procedures are designed to reduce ICI, which incurs high computational complexity and results in error propagation.

  In contrast to CS theory that reconstructs each sparse signal individually, the distributed compressive sensing (DCS) proposed in \cite{DCS2005} aims to jointly reconstruct a collection of sparse signals by exploiting their joint sparsity. In  \cite{DCS2005}, a DCS simultaneous orthogonal matching pursuit (DCS-SOMP) algorithm is proposed to reconstruct jointly sparse signals. It is shown in \cite{dcsVScs,scs,pengchengTcom} that DCS-based methods achieve higher estimation accuracy than CS-based methods. In \cite{pengchengTcom}, a novel DCS based channel estimation scheme is proposed to track the DS channel with a large Doppler shift, which introduces an ICI-free structure without additional iterative operations. However, \cite{pengchengTcom} ignores the temporal correlation of sparse channels and merely exploits the sparse characteristic of channel parameters within a single OFDM symbol.

    Further studies \cite{TDSmultiple,Linglong2013} have shown the temporal correlation of practical time-varying channels: although the path gains will change over adjacent OFDM symbols, the path delays may remain relatively unchanged. This observation motivates us to seek effective joint multi-symbol channel estimation methods to enhance the estimation precision over a DS channel. Some works regarding joint multi-symbol channel estimation methods have been reported in literature \cite{multisym, zijian2006, zijian2011}. However, these works are not based on DCS theory which can improve spectral efficiency and estimation accuracy. To our best knowledge, little has been done about applying DCS theory to joint multi-symbol channel estimations over a DS channel.

  In this paper, we propose a novel structured DCS (SDCS) based joint multi-symbol channel estimation scheme. To be specific, in order to reduce the number of unknown channel coefficients, we utilize the complex exponential basis expansion model (CE-BEM) to model the time variation of a DS channel within multiple OFDM symbols. Then, by exploiting the sparsity in the delay domain and designing special pilot pattern within multiple OFDM symbols, we are able to decouple jointly sparse CE-BEM coefficient vectors, leading to a special jointly sparse block structure in the aggregate coefficient vectors. To obtain a channel estimator consistent with this joint-block-sparse model, a block-based SOMP (BSOMP) algorithm derived from the classical SOMP algorithm is proposed to jointly recover the channel parameters. Based on the exploitation of the structural property in the model, we can expect the channel estimation performance to be significantly improved.

  In order to reduce the modeling error in the CE-BEM when modeling the DS channel, we carry out smoothing treatments of already estimated channel taps via piecewise linear approximation. Jeon and Chang have assumed a linear model for channel variations under the condition of low Doppler shift in \cite{linear1999}. On the basis of \cite{linear1999}, \cite{Linear2005} has proved piecewise linear approximation is a good estimate of channel time-variations even for normalized Doppler of up to 0.2. It is also found that in \cite{linewater1}, \cite{linewater2}, a piecewise linear model is used to approximate time-varying underwater acoustic channels. In this paper, we propose two novel smoothing treatment methods within a single OFDM symbol and within multiple OFDM symbols, respectively. Both of our methods are based on a piecewise linear approximation for a DS channel. Simulation results demonstrate that the proposed smoothing treatment can significantly improve the channel estimation performance.

   The main contributions of this paper lie in two aspects. One is that exploiting the temporal correlations of a DS channel, we propose a novel SDCS based joint multi-symbol estimation model, and a novel BSOMP algorithm is proposed to solve the model. The other is that we propose two novel smoothing methods via piecewise linear approximation. Our simulation results show that when dealing with the joint multi-symbol channel estimation model, the proposed SDCS-based scheme can achieve higher estimation accuracy than the CS-based and DCS-based schemes. And the proposed smoothing treatment scheme can significantly improve the channel estimation accuracy. Further, it is shown that the proposed  joint multi-symbol channel estimation scheme is superior to the conventional single-symbol channel estimation scheme \cite{pengchengTcom} in terms of both estimation accuracy and spectrum efficiency.

 The remainder of this paper is organized as follows. Section \uppercase\expandafter{\romannumeral 2} introduces an OFDM system model over a DS channel and the CE-BEM. Section \uppercase\expandafter{\romannumeral 3} describes formulation of SDCS-based channel estimation model. Section \uppercase\expandafter{\romannumeral 4} describes the proposed BSOMP algorithm and the process of smoothing treatment. In section \uppercase\expandafter{\romannumeral 5}, simulation results are provided to demonstrate the superior performance of our proposed scheme. Finally in Section \uppercase\expandafter{\romannumeral 6}, some concluding remarks are given.

 $Notations:$
   For a given matrix $\mathbf A$, ${\mathbf A^{-1} }$, ${\mathbf A^{\dagger} }$, ${\mathbf A^T}$ and ${\mathbf A^H }$ denote its inverse, pseudo inverse, transpose and conjugate transpose, respectively. ${\left\| \mathbf A \right\|_2}$ denotes the Frobenius-norm of $\mathbf A$. $[\mathbf A]_{k,n}$, $[\mathbf A]_{\mathcal P,\mathcal L}$ and $[\mathbf A]_{\mathcal P}$ denote $(k,n)$-th entry of a matrix $\mathbf A$, a submatrix of $\mathbf A$ with row indices $\mathcal P$ and column indices $\mathcal L$, and a submatrix of $\mathbf A$ with row indices $\mathcal P$ and all columns, respectively. $\bm{\mathcal D}\{\mathbf A_0,\ldots,\mathbf A_{N-1}\}$ denotes a block-wise diagonal matrix with the matrices $\mathbf A_0,\ldots,\mathbf A_{N-1}$ on the diagonal. For a given vector $\mathbf a$, ${\left\| \mathbf a \right\|_p}$ (subject to $p\geq 1$) denotes the $p$-norm of vector $\mathbf a$, and $\bm{\mathcal D}\{\mathbf a\}$ denotes a diagonal matrix with $\mathbf a$ on its main diagonal. $\mathbb R^{M\times N}$ and $\mathbb C^{M\times N}$ represent the set of $M\times N$ matrices in real field and complex field, respectively. $\otimes$ represents the Kronecker product. $\mathbf I_N$ stands for an $N\times N$ identity matrix, $\mathbf 1_N$ for the $N\times 1$ column vector of all ones, and $\mathbf F_N$ for a $N$-point normalized discrete Fourier transform (DFT) matrix with $[\mathbf F_N]_{n,m} = 1/\sqrt{N}e^{\frac{-j2\pi nm}{N}}(n,m\in [0,N-1])$. The cardinality of the set $\mathcal S$ is denoted by $|\mathcal S|$. $E(z)$ represents the average of $z$.

\section{System model}

 Here, we first introduce the fundamental model of OFDM systems over a DS channel. Then, we describe the CE-BEM within a single OFDM symbol and extend the model to multiple OFDM symbols.

\subsection{OFDM System Model over a DS Channel}
  We consider an OFDM transmission system with \emph{N} subcarriers, and use $h_{n,l}$ to denote the channel gain of the $l$-th $(l\in [0,L-1])$ discrete path of the CIR at time $n$. The transmit signal of the \emph{j}-th OFDM symbol is denoted as $\mathbf{X}^{(j)}=[X^{(j)}(0),\ldots,X^{(j)}(N-1)]$, for $j\in [0,J-1]$. Let us use $[\mathbf{X}^{(j)}]_{\mathcal{P}}$ to denote the pilots of the $j$-th OFDM symbol, where $\mathcal{P}$ ($|\mathcal{P}|=P$) is the set of pilot subcarrier indices, and $[\mathbf{X}^{(j)}]_{\mathcal{D}}$ to denote the corresponding data, where $\mathcal{D}$ ($|\mathcal{D}|=N-P$) is the set of data subcarriers indices.

   Once performing inverse DFT (IDFT) on $\mathbf{X}^{(j)}$, we can express the time-domain modulated signal as $\mathbf{x}^{(j)} = \mathbf{F}_{N}^{H}\mathbf{X}^{(j)}$. In order to avoid the ISI resulting from multipath delay spreads, the time-domain signal is concatenated by a cyclic prefix (CP) with length $L_{CP}~(L_{CP}\geq L)$. Finally, the symbol streams are converted from a parallel to a serial form and transmitted through a DS channel.

  At the receiver side, after removing the CP, we demodulate the remaining samples by N point DFT matrix $\mathbf{F}_{N}$. The received signal of the \emph{j}-th OFDM symbol can be expressed as
  \begin{eqnarray}
    \mathbf{Y}^{(j)} = \underbrace{{\mathbf{F}_N}{\mathbf{H}_T^{(j)}}\mathbf{F}_N^H}_{\mathbf{H}_F^{(j)}}\mathbf X^{(j)} + \mathbf{W}^{(j)},
  \end{eqnarray}
  where $\mathbf{H}_T^{(j)}$ is the $N\times N$ matrix in time-domain including the effects of concatenating and removing the CP, $\mathbf{H}_F^{(j)}\in \mathbb C^{N\times N}$ represents the corresponding frequency-domain channel matrix, and $\mathbf{W}^{(j)}\in \mathbb C^{N\times 1}$ denotes the additive noise. To be specific, the $(p,q)$-th $(p,q \in [0,N - 1])$ entry of $\mathbf{H}_T^{(j)}$ can be expressed as
  \begin{eqnarray}\label{HT}
[\mathbf{H}_T^{(j)}]_{p,q} = h_{j(N+L_{CP})+L_{CP} + p,mod(p - q,N)},
  \end{eqnarray}
  where $mod(a,b)$ stands for the remainder of $a$ divided by $b$.

  Clearly, if the channel is time-invariant, $\mathbf H_{T}^{(j)}$ will be a circular matrix and as a result $\mathbf{H}_{F}^{(j)}$ will be a diagonal matrix. While in a DS channel, $\mathbf{H}_{T}^{(j)}$ exhibits pseudo-circular structure, which results in a full matrix $\mathbf H_{F}^{(j)}$ instead of a diagonal one and thus induces ICI.

  For channel estimation during \emph{J} consecutive OFDM symbols, $JNL$ channel coefficients of $h_{n,l}$ need to be estimated. Thus we should allocate pilot subcarriers on the order of $JNL$, which is very large. In the following subsection, we will introduce the CE-BEM to reduce the total number of coefficients to be estimated.

\subsection{CE-BEM in the Time Domain}
  In this subsection, we will try to model the time-variation of a DS channel by using the CE-BEM due to the temporal $(n)$ variation of $h_{n,l}$ is usually rather smooth. Let us define the \emph{l}-th $(l\in [0,L-1])$ channel tap related to the \emph{j}-th OFDM symbol as $\mathbf{h}_l^{(j)} \buildrel \Delta \over = {\left( {h_{j(N + {L_{CP}}) + {L_{CP}},l}, \ldots ,h_{(j+1)(N + {L_{CP}})- 1,l}} \right)^T} \in \mathbb C^{N\times 1}$, which can be expressed as
 \begin{eqnarray}\label{ce-bem}
 \mathbf{h}_{l}^{(j)}= \left( {{\mathbf{b}_0}~ \cdots ~ {\mathbf{b}_{Q - 1}}} \right)\left( {\begin{array}{*{20}{c}}
{{c^{\left( j \right)}}\left[ {0,l} \right]}\\
 \vdots \\
{{c^{\left( j \right)}}\left[ {Q - 1,l} \right]}
\end{array}} \right) +\bm{\xi} _l^{\left( j \right)},
  \end{eqnarray}
  where $Q~(Q\ll N)$ denotes the BEM order,  $\mathbf{b}_{q}\in \mathbb C^{N\times 1}(q\in [0,Q-1])$ is the orthonormal basis function, $c^{(j)}[q,l]$ represents the corresponding BEM coefficient related to the $j$-th OFDM symbol, and $\bm{\xi} _l^{\left( j \right)}\in \mathbb C^{N\times 1}$ denotes the BEM modeling error. Note that $\bm{\xi} _l^{\left( j \right)}$ will be dealt with by a piecewise linear smoothing treatment proposed in Section \uppercase\expandafter{\romannumeral 4}. To be specific, the CE-BEM basis functions are complex exponential with a period of $N$, and the $q$-th basis function $\mathbf{b}_{q}$ can be expressed as
  \begin{eqnarray}
{{\rm{\mathbf{b}}}_q} = {\left( {1, \ldots ,{e^{j\frac{{2\pi }}{N}n\left( {q - \frac{{Q - 1}}{2}} \right)}}, \ldots ,{e^{j\frac{{2\pi }}{N}\left( {N - 1} \right)\left( {q - \frac{{Q - 1}}{2}} \right)}}} \right)^T}.
  \end{eqnarray}
  The CE-BEM is able to make the frequency-domain channel matrix $\mathbf{H}_{F}^{(j)}$ strictly banded \cite{pengchengTcom}. Note that to exploit the symmetrical property in the sequel, we assume that $Q$ is an odd number.

  We further define $\mathbf{c}_q^{\left( j \right)} \buildrel \Delta \over = {\left( {{c^{\left( j \right)}}[q,0], \ldots ,{c^{\left( j \right)}}[q,L - 1]} \right)^T}\in \mathbb C^{L\times 1}$. Then the time-domain channel matrix $\mathbf{H}_T^{(j)}$ given in (2) can be illustrated in terms of the CE-BEM as
  \begin{equation}
    \mathbf{H}_T^{(j)} = \sum\limits_{q = 0}^{Q-1} {\bm{\mathcal{D}}\left\{ {{\mathbf{b}_q}} \right\}\mathbf{F}_N^H\mathbf{\bm{\mathcal{D}}}\left\{ {{{\mathbf{V}}_L}{\mathbf{c}}_q^{(j)}} \right\}{\mathbf F_N}}+\bm{\xi}^{(j)},
  \end{equation}
   where $\mathbf{V}_{L}\in \mathbb C^{N\times L}$ denotes the submatrix that extracts the first $L$ columns of $\sqrt{N}\mathbf{F}_{N}$ \cite{zijian2011}, which can be written as
  \begin{equation}
{\mathbf V_L} = {\left( {\begin{array}{*{20}{c}}
1&1& \cdots &1\\
1&w& \cdots &{{w^{L - 1}}}\\
 \vdots & \vdots & \vdots & \vdots \\
1&{{w^{N - 1}}}& \cdots &{{w^{\left( {N - 1} \right)\left( {L - 1} \right)}}}
\end{array}} \right)_{N \times L}},
  \end{equation}
  with $w \buildrel \Delta \over = \exp \left( { - i\frac{{2\pi }}{N}} \right)$.
  Substituting (5) into (1), we can obtain the received signal of the \emph{j}-th OFDM symbol in terms of the CE-BEM as
  \begin{equation}
    \mathbf{Y}^{(j)} = \underbrace{\sum\limits_{q = 0}^{Q-1} {\mathbf{I}_N^{\langle q-\frac{Q-1}{2} \rangle}\mathbf{\bm{\mathcal{D}}}\left\{ {{{\mathbf{V}_L}\mathbf{c}_q^{(j)}}} \right\}}}_{\mathbf{H}_F^{(j)}} {\mathbf{X}^{(j)}} +\mathbf{Z}^{(j)},
  \end{equation}
  where $\mathbf{I}_N^{\langle q \rangle}\in \mathbb C^{N\times N}$ denotes a permutation matrix obtained from $\mathbf{I}_N$ by shifting its column circularly $|q|$-times to the left if $q>0$ and to the right otherwise, and $\mathbf{Z}^{(j)}\in \mathbb C^{N\times 1}$ includes the additive noise and the CE-BEM modeling error.

  Accordingly, we can express the received $J$ consecutive OFDM symbols as
  \begin{equation}\label{eq_mul}
\mathbf{Y} = \left[ {\begin{array}{*{20}{c}}
\mathbf{H}_F^{(0)}&{}&{}\\
{}& \ddots &{}\\
{}&{}&\mathbf{H}_F^{(J-1)}
\end{array}} \right]\mathbf{X}+\mathbf Z,
\end{equation}
where $\mathbf{Y}  \buildrel \Delta \over = {\left( {{\mathbf Y^{(0)}}^T, \ldots ,{\mathbf Y^{(J - 1)}}^T} \right)^T}\in \mathbb C^{JN\times 1}$ and $\mathbf{X}  \buildrel \Delta \over = {\left( {{\mathbf X^{(0)}}^T, \ldots ,{\mathbf X^{(J - 1)}}^T} \right)^T}\in \mathbb R^{JN\times 1}$ denote the received and transmitted subcarriers during $J$ consecutive OFDM symbols, respectively, $\mathbf{Z}  = {\left( {{\mathbf Z^{(0)}}^T, \ldots ,{\mathbf Z^{(J - 1)}}^T} \right)^T}\in \mathbb C^{JN\times 1}$ represents the total error and noise, and $\mathbf{H}_F^{(j)}~(j\in [0,J-1])$ is expressed in (7). Note that the CE-BEM basis functions $\{\mathbf b_q\}_{q=0}^{Q-1}$ are common for each OFDM symbol, but the CE-BEM coefficient vectors $\{\mathbf c_q^{(j)}\}_{q=0}^{Q-1}$ are not.

Therefore, instead of estimating numerous channel taps $\{\mathbf{h}_l^{(j)}\}_{l=0}^{L-1}$, we turn to identify the CE-BEM coefficient vectors $\{\bm c_q^{(j)}\}_{q=0}^{Q-1}$. Obviously, the CE-BEM is able to dramatically reduce the total number of unknown coefficients within $J$ consecutive OFDM symbols from $JNL$ to $JQL$ with $Q\ll N$.

 \section{Proposed Channel Estimation Scheme}
  In this section, we first briefly introduce some basic CS and DCS theories. Then, we give a detailed description about the joint sparsity of the CE-BEM coefficients within multiple OFDM symbols. Next, we design a special pilot pattern and thus transform original channel estimation problem into a SDCS form.


\subsection{CS theory}
  CS is a revolutionary technique to reconstruct a sparse signal from an undetermined model. Consider $\mathbf{Y}=\mathbf{\Phi}\bm{\theta}+\bm{\eta}$, where $\mathbf{\Phi}$ is an $G\times L$ matrix with $G<L$, $\bm{\theta} \in \mathbb{C}^{L}$ is an unknown signal vector, \textbf{Y} $\in \mathbb{C}^{G}$ represents the observed vector, and $\bm \eta \in \mathbb C^{G}$ denotes a noise vector. The goal of CS is to reconstruct $\bm \theta$ correctly based on the knowledge of $\mathbf Y$ and $\mathbf{\Phi}$. Fundamental researches \cite{RIP1}, \cite{RIP2} indicate that if $\mathbf\Phi$ satisfies the restricted isometry property (RIP) and $\bm \theta$ has merely $K~(K\ll L)$ nonzero values, $\bm \theta$ can be reconstructed correctly with CS reconstruction methods such as the basis pursuit (BP) and the orthogonal matching pursuit (OMP) even under an undetermined condition.

  However, it incurs tremendous computational complexity to verify that $\mathbf \Phi$ satisfies the RIP. To simplify the calculation, we consider the mutual coherence property (MCP) as alternative property, which has been widely adopted in the literature. The mutual coherence of $\mathbf\Phi$ can be expressed as
  \begin{eqnarray}
    \mu(\mathbf{\Phi})= \mathop {\max }\limits_{1 \le i \ne j \le L} \frac{{\left| {\left\langle {{\bm \phi _i},{\bm \phi _j}} \right\rangle } \right|}}{{{{\left\| {{\bm \phi _i}} \right\|}_2}{{\left\| {{\bm \phi _j}} \right\|}_2}}},
  \end{eqnarray}
  where $\bm {\phi}_i$ and $\bm {\phi}_j$ are the two arbitrary columns of $\mathbf{\Phi}$. According to \cite{scs}, the smaller $\mu(\mathbf{\Phi})$ is, the more accurately $\bm \theta$ will be recovered.

\subsection{DCS theory}
  The DCS theory extends the CS theory to recover a set of multiple correlated signals. Instead of reconstructing a single sparse signal alone, the objective of DCS is to reconstruct a collection of jointly sparse signals from the same measurement matrix $\mathbf{\Phi}$ satisfying MCP. Let us consider a set of $Q$ undetermined problems
  \begin{eqnarray}\label{Eq_CS2}
    \mathbf Y_{q}=\mathbf \Phi \bm\theta_{q}+\bm \eta_{q},  ~~~q\in \{0,1,\ldots,Q-1\},
  \end{eqnarray}
  where $\mathbf{Y}_{q}$ $\in \mathbb C^{G}$, $\mathbf \Phi \in \mathbb C^{G\times L}$, $\bm \eta_{q} \in \mathbb C^{G}$ and $\bm \theta_{q} \in \mathbb C^{L}$. Here, each vector $\bm \theta_{q}$ are jointly sparse, i.e., not only does each vector $\bm \theta_{q}$ have $K$ nonzero entries, but also the nonzero entries in all $\bm \theta_{q}$ occur in the same positions.

  Let us write (\ref{Eq_CS2}) in a combined form as
  \begin{eqnarray}
    \bar{\mathbf{Y}}=\mathbf\Phi \bar{\bm \theta}+\bar{\bm\eta},
  \end{eqnarray}
  where $\bar{\mathbf{Y}}=(\mathbf{Y}_{0},\ldots,\mathbf{Y}_{Q-1}) \in \mathbb C^{G\times Q}$, $\bar{\bm\theta}=(\bm\theta_{0},\ldots,\bm\theta_{Q-1}) \in \mathbb C^{L\times Q}$, and $\bar{\bm\eta}=(\bm\eta_{0},\ldots,\bm\eta_{Q-1}) \in \mathbb C^{G\times Q}$. To recover jointly sparse signals, a DCS-SOMP algorithm was proposed in \cite{DCS2005}. Obviously, we can carry out CS theory to reconstruct each $\bm \theta_{q}$ from $\mathbf Y_{q}$ individually. However, it has been verified in \cite{scs} that under the condition of the same number of samples, DCS outperforms CS notably in terms of recovery accuracy. This advantage is owing to the fact that the joint processing in DCS can yield higher possibility of searching the correct location of nonzero values.


 \subsection{Sparsity of the CE-BEM Coefficient Vectors within Multiple OFDM Symbols}
    In a broadband system with a large bandwidth $B$ and a small number of propagation paths, the delay interspacings are usually larger than the delay resolution bin of width $\Delta\tau=1/B$. Not every delay bin of size $\Delta\tau$ contains a physical path. Thus the delay domain exhibits sparsity \cite{learningDS2008}. To explore the sparsity of a DS channel in the delay domain, we introduce the definition of $K$-sparse channel based on \cite{CCSwaheed}.


    \emph{Definition~1}: For a fixed $n$, suppose that $\mathcal L=\{l:|h[n,l]|>\varepsilon\}$ denotes the set of indices of dominant channel coefficients of a wireless channel for some appropriately chosen $\varepsilon$. We say that the channel is effectively $K$-sparse in the delay domain if it satisfies $K=|\mathcal L|\ll L$, where $L$ is the maximum number of resolvable paths.


    For a DS channel, practical wireless channels exhibit temporal correlations. The path delays usually vary much slower than the path gains \cite{multisym2000}. This is because the duration $T_{delay}$ for the path delay variation is inversely proportional to the signal bandwidth $f_s$, while the coherence time of time-varying path gains $T_{gain}$ is inversely proportional to the system's carrier frequency $f_c$ \cite{Linglong2013}. Since we have $f_s\ll f_c$ for a practical wireless system, we obtain $T_{gain}\ll T_{delay}$. In \cite{TDSmultiple}, we observe that during several consecutive OFDM symbols, although the path gains will be quite different, the path delays typically remain unchanged.
 \begin{figure}[!t]
    \centering
    \includegraphics[width=9cm]{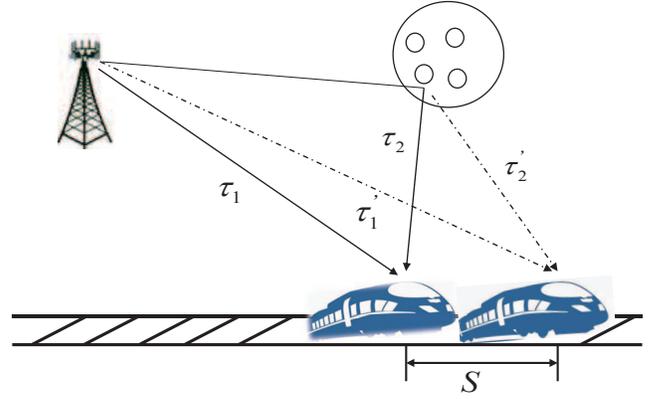}
    \caption{Illustration for multipath signal transmission}.
    \label{fig.1}
    \end{figure}

    Let us now elaborate this formally. In Fig.~\ref{fig.1}, during $J$ consecutive OFDM symbols, we calculate the displacement of the vehicle as $S=JT_s\left(N+L_{CP}\right)v$, where $T_s$ is the sampling period and $v$ is the velocity of the vehicle. It is easy to show that the maximum variation of all path delays $\Delta {\tau _{\max }} \leq S/c$, where $c$ is the speed of light. As a result, we have
    \begin{equation}
    \Delta {\tau _{\max }} \leq J{T_s}\left( {N + {L_{CP}}} \right)v/c.
    \end{equation}
    For a fixed $T_s$, once it satisfies $\Delta {\tau _{\max }}/{T_s}< 0.01$, the maximum variation of all path delays is much smaller than the sampling period $T_s$ such that we could assume the path delays remain relatively unchanged.

    Consequently, when the number of consecutive OFDM symbols $J$ subjects to
    \begin{equation}
   J < \frac{{0.01c}}{{\left( {N + {L_{CP}}} \right)v}},
    \end{equation}
    we could assume that the positions of nonzero entries in CIR within $J$ consecutive symbols remain unchanged. Thus we obtain $h_{n,l}=0~(n\in [0,J(N+L_{CP})-1])$ for $l\notin \mathcal L$, where $\mathcal L$ denotes the aggregate dominant paths described in \emph{Definition~1}. Then we have $\mathbf h_l^{(j)}=0~(j\in [0,J-1])$ for $l\notin \mathcal L$. Further, based on (3), it is easy to show that
    \begin{equation}
      c^{(j)}[0,l]=\cdots =c^{(j)}[Q-1,l]=0~(l\notin \mathcal L)
    \end{equation}
    due to $(c^{(j)}[0,l],\cdots,c^{(j)}[Q-1,l])^T=(\mathbf b_0,\ldots,\mathbf b_{Q-1})^{\dagger}\mathbf h_l$ regardless of the modeling error. Consequently, $\mathbf{c}_q^{(j)}$ will be a sparse vector with a sparsity of $K$ and all $\mathbf{c}_q^{(j)}~(j\in [0,J-1],q\in [0,Q-1])$ share the common locations of nonzero values, i.e., $\{\mathbf{c}_0^{(0)},\ldots,\mathbf{c}_{Q-1}^{(0)},\ldots,\mathbf{c}_0^{(J-1)},\ldots,\mathbf{c}_{Q-1}^{(J-1)}\}$ are jointly sparse.

  \subsection{The SDCS Formulation}
  In this subsection, we will extend it the same idea in \cite{pengchengTcom} to design sparse pilot pattern but seek optimal pilot placement among $J$ consecutive OFDM symbols, which is related to the CE-BEM order $Q$. Then an ICI free structure is obtained and finally the channel estimation problem is formulated into an SCDS framework.

   We denote the total number of pilot subcarriers within $J$ OFDM symbols as $P$, and the corresponding pilot indices as $\mathcal P$. The pilot subcarriers are grouped in $G~(K<G\ll JL)$ clusters. Each cluster includes one value pilot and $(2Q-2)$ guard pilots. The value pilot index set $\mathcal P_{val}$ ($|\mathcal P_{val}|=G$) is expressed as
        \begin{equation}
          \mathcal P_{val}=\{p_0,\ldots,p_{G-1}\},
        \end{equation}
         where $0\leq p_0<\cdots<p_{G-1}\leq JN-1$. And the guard pilot index $\mathcal P_{guard}$ ($|\mathcal P_{guard}|=(2Q-2)G$) is given by
        \begin{equation}
          \mathcal P_{guard}=\cup\{k-Q+1,\ldots,k-1,k+1,\ldots,k+Q-1\},
        \end{equation}
        where ${k\in \mathcal P_{val}}$. We set the elements in the value pilot subcarriers $\mathbf P_{val}\in \mathbb C^{G}$ with constant amplitude and the elements in the guard pilot subcarriers $\mathbf P_{guard}$ as zero.

  It is clear that $|\mathcal P_{val}|+|\mathcal P_{guard}|=(2Q-1)G=P$ and $\mathcal P_{val}\cup\mathcal P_{guard}=\mathcal P$. Note that we must have $|p_i-p_j|\geq 2Q-1, i\neq j$, to prevent the locations of the value pilot subcarriers and the guard pilot subcarriers overlapping.

  Next, we re-divide pilot indices $\mathcal P$ into $Q$ subsets, denoted as
    \begin{equation}
    \left \{{\begin{array}{*{20}{c}}
    {\begin{array}{*{20}{c}}
    {{\mathcal P_0} = {\mathcal P_{val}} - \frac{{Q - 1}}{2}}~~\\
     \vdots
    \end{array}}\\
    {{\mathcal P_{\frac{{Q - 1}}{2}}} = {\mathcal P_{val}}}~~~~~~~\\
     \vdots \\
    {{\mathcal P_{Q - 1}} = {\mathcal P_{val}} + \frac{{Q - 1}}{2}},
    \end{array}}\right.
  \end{equation}
   where $\mathcal P_{val}-\frac{Q-1}{2}$ stands for a new set with all elements in $\mathcal P_{val}$ subtract $\frac{Q-1}{2}$. Such an arrangement of the pilot pattern $\mathcal P$ with $Q=3$ is depicted in Fig.~2.
     \begin{figure}[!th]
    \centering
    \includegraphics[width=9cm]{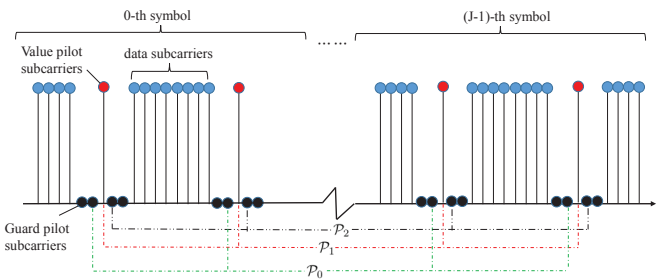}
    \caption{The pilot pattern ($Q=3$)}.
    \label{fig.2}
    \end{figure}

   Based on the designed sparse pilot pattern and properties of the CE-BEM, the estimation of $JQ$ sparse CE-BEM coefficient vectors $\{\bm c_q^{(j)}\}_{q=0}^{Q-1}$ could be decoupled from (\ref{eq_mul}) by $Q$ separate equations without ICI as

  \begin{equation}\label{EQ_mulmodel}
   \left\{{ \begin{array}{*{20}{c}}
    {\begin{array}{*{20}{c}}
    {{{\left[ \mathbf Y \right]}_{{\mathcal P_0}}} = \mathbf \Psi{{\left[ {{\mathbf I_J} \otimes  {{\mathbf V_L}} } \right]}_{{\mathcal P_{val}}}}{\left( {\begin{array}{*{20}{c}}
    {\mathbf c_{0}^{(0)}}\\
     \vdots \\
    {\mathbf c_{0}^{(J-1)}}
    \end{array}} \right)} + {\mathbf Z_0}}~~~~~~\\
     \vdots
    \end{array}}\\
    {{{\left[ \mathbf Y \right]}_{{\mathcal P_{\frac{{Q - 1}}{2}}}}} = \mathbf \Psi{{\left[ {\mathbf I_J} \otimes  {{\mathbf V_L}}  \right]}_{{\mathcal P_{val}}}}{\left( {\begin{array}{*{20}{c}}
    {\mathbf c_{\frac{{Q - 1}}{2}}^{(0)}}\\
     \vdots \\
    {\mathbf c_{\frac{{Q - 1}}{2}}^{(J-1)}}
    \end{array}} \right)}  + {\mathbf Z_{\frac{{Q - 1}}{2}}}}\\
     \vdots \\
    {{{\left[ \mathbf Y \right]}_{{\mathcal P_{Q - 1}}}} = \mathbf \Psi{{\left[ {\mathbf I_J} \otimes {{\mathbf V_L}} \right]}_{{\mathcal P_{val}}}}{\left( {\begin{array}{*{20}{c}}
    {\mathbf c_{Q-1}^{(0)}}\\
     \vdots \\
    {\mathbf c_{Q-1}^{(J-1)}}
    \end{array}} \right)}  + {\mathbf Z_{Q - 1}}},
    \end{array}} \right.
  \end{equation}
   where $\mathbf \Psi=\bm{\mathcal D}\{\mathbf P_{val}\}$ denotes a diagonal matrix with the value pilot subcarriers on its diagonal, $[\mathbf Y]_{\mathcal P_q}\in \mathbb C^{G\times 1}$ represents the subset of received $J$ consecutive OFDM subcarriers $\mathbf Y$ corresponding to $\mathcal P_q$, $\mathbf V_L$ is given in (6), and $\mathbf Z_q\in \mathbb C^{G\times 1}$ includes the noise and the modeling error. (Please refer to Appendix for the complete proof of (\ref{EQ_mulmodel}).)

   We further define $\mathbf c_q' \buildrel \Delta \over =((\mathbf{c}_q^{(0)})^T,\ldots,(\mathbf{c}_{q}^{(J-1)})^T)^T\in \mathbb C^{JL\times 1}$. Since we have verified in section C that the aggregate coefficient vectors in~$\{\mathbf{c}_0^{(0)},\ldots,\mathbf{c}_{Q-1}^{(0)},\ldots,\mathbf{c}_0^{(J-1)},\ldots,\mathbf{c}_{Q-1}^{(J-1)}\}$~are jointly sparse, we obtain $\{\mathbf c_q'\}_{q=0}^{Q-1}$ are also jointly sparse. In addition, each equation in (\ref{EQ_mulmodel}) shares the same measurement matrix ${{\mathbf \Psi}{{\left( {{\mathbf I_J} \otimes \mathbf V_L} \right)}_{{\mathcal P_{val}}}}}$.   Consequently, we are able to estimate $\{\mathbf c_q'\}_{q=0}^{Q-1}$ based on DCS theory.

   However, it can be observed that the coefficient vectors $\{\mathbf c_q'\}_{q=0}^{Q-1}$ have the inherent structured sparsity, which motivates us to apply the theory of SDCS instead of the conventional DCS theory to estimate sparse coefficient vectors. Let us rearrange the elements of the vector $\mathbf c_q'$ as
   \begin{equation}\label{s1}
    {\mathbf s_q} = {\left( {{{\left( {\mathbf s_q^0} \right)}^T}, \ldots ,{{\left( {\mathbf s_q^{L - 1}} \right)}^T}} \right)^T\in \mathbb C^{JL\times 1}},
  \end{equation}
  with
  \begin{equation}\label{s2}
    \mathbf s_q^l = (\mathbf c_q'(l),\ldots,\mathbf c_q'((J-1)L+l)^T\in \mathbb C^{J\times 1}.
  \end{equation}

   Then the system model of (\ref{EQ_mulmodel}) can be reobtained as
  \begin{equation}\label{EQ_lastmodel}
    \left \{{\begin{array}{*{20}{c}}
    {\begin{array}{*{20}{c}}
    {{{\left[ \mathbf Y \right]}_{{\mathcal P_0}}} = \mathbf \Phi {\mathbf s_0} + {\mathbf Z_0}}~~~~~~~~~~\\
     \vdots
    \end{array}}\\
    {{{\left[ \mathbf Y \right]}_{{\mathcal P_{\frac{{Q - 1}}{2}}}}} = \mathbf \Phi {\mathbf s_{\frac{{Q - 1}}{2}}} + {\mathbf Z_{\frac{{Q - 1}}{2}}}}\\
     \vdots \\
    {{{\left[ \mathbf Y \right]}_{{\mathcal P_{Q - 1}}}} = \mathbf \Phi {\mathbf s_{Q - 1}} + {\mathbf Z_{Q - 1}}},
    \end{array}}\right.
  \end{equation}
  where the new measurement matrix $\mathbf{\Phi}  = ({\mathbf{\Phi} _0}, \ldots, {\mathbf{\Phi} _{L - 1}})\in \mathbb C^{G\times JL}$ with $\mathbf{\Phi} _l\in \mathbb C^{G\times J}$ expressed as
  \begin{equation}
    {\mathbf \Phi _l} = {\left[ {{\mathbf \Psi}{{\left( {{\mathbf I_J} \otimes \mathbf V_L} \right)}_{{\mathcal P_{val}}}}} \right]_{l:L:(J - 1)L + l}}.
  \end{equation}
  Here, $[\mathbf A]_{l:L:(J-1)L+l}$ denotes a submatrix extracting the columns of $\mathbf A$ according to the indices $\{l,L+l,\ldots,(J-1)L+l\}$.

  Thanks to the inherent structured sparsity of $\{\mathbf c_q'\}_{q=0}^{Q-1}$, the rearranged coefficient vectors $\{\mathbf s_q\}_{q=0}^{Q-1}$ exhibit block sparsity as well as joint sparsity, leading to a special jointly sparse block structure in the system model. In addition, each equation in (\ref{EQ_lastmodel}) shares the same measurement matrix $\mathbf \Phi$. We write (\ref{EQ_lastmodel}) in a more compact form and obtain the SDCS model as
  \begin{equation}\label{eq_lastcompact}
    ([\mathbf{Y}]_{\mathcal{P}_{0}},\ldots,[\mathbf{Y}]_{\mathcal{P}_{Q-1}})=\mathbf \Phi (\mathbf s_{0},\ldots,\mathbf s_{Q-1})+\mathbf Z.
  \end{equation}

  Now there are two remaining major issues, listed as below
  \begin{enumerate}
    \item The common measurement matrix $\mathbf \Phi$ needs to meet MCP, which motives us to seek optimal pilot placement to make $\mu(\mathbf{\Phi})$ as small as possible.
    \item We need to propose an efficient algorithm corresponding to the SDCS model.
  \end{enumerate}

  To solve the first problem, we formulate the optimization problem as
  \begin{equation}\label{EQ_u}
\begin{array}{*{20}{c}}
{\mathop {\min }\limits_{{\mathcal P_{\frac{{Q - 1}}{2}}}} {\kern 1pt} {\kern 1pt} {\kern 1pt} {\kern 1pt} {\kern 1pt} {\kern 1pt} {\kern 1pt} {\kern 1pt} {\kern 1pt} {\kern 1pt} {\kern 1pt} {\kern 1pt} \mu \left( \mathbf \Phi  \right){\kern 1pt} {\kern 1pt} {\kern 1pt} {\kern 1pt} {\kern 1pt} {\kern 1pt} {\kern 1pt} {\kern 1pt} {\kern 1pt} {\kern 1pt} {\kern 1pt} {\kern 1pt} {\kern 1pt} {\kern 1pt} {\kern 1pt} {\kern 1pt} {\kern 1pt} {\kern 1pt} {\kern 1pt} {\kern 1pt} {\kern 1pt} {\kern 1pt} {\kern 1pt} {\kern 1pt} {\kern 1pt} {\kern 1pt} {\kern 1pt} {\kern 1pt} {\kern 1pt} {\kern 1pt} {\kern 1pt} {\kern 1pt} {\kern 1pt} {\kern 1pt} {\kern 1pt} {\kern 1pt} {\kern 1pt} {\kern 1pt} {\kern 1pt} {\kern 1pt} {\kern 1pt} {\kern 1pt} {\kern 1pt} {\kern 1pt} {\kern 1pt} {\kern 1pt} {\kern 1pt} {\kern 1pt} {\kern 1pt} {\kern 1pt} {\kern 1pt} {\kern 1pt} {\kern 1pt} {\kern 1pt} {\kern 1pt} {\kern 1pt} {\kern 1pt} {\kern 1pt} {\kern 1pt} {\kern 1pt} {\kern 1pt} {\kern 1pt} {\kern 1pt} {\kern 1pt} {\kern 1pt} {\kern 1pt} {\kern 1pt} {\kern 1pt} {\kern 1pt} {\kern 1pt} {\kern 1pt} {\kern 1pt} {\kern 1pt} {\kern 1pt} {\kern 1pt} {\kern 1pt} {\kern 1pt} {\kern 1pt} {\kern 1pt} {\kern 1pt} {\kern 1pt} {\kern 1pt} {\kern 1pt} {\kern 1pt} {\kern 1pt} {\kern 1pt} {\kern 1pt} {\kern 1pt} {\kern 1pt} {\kern 1pt} {\kern 1pt} {\kern 1pt} {\kern 1pt} {\kern 1pt} {\kern 1pt} {\kern 1pt} {\kern 1pt} }\\
{s.t.{\kern 1pt} {\kern 1pt} {\kern 1pt} {\kern 1pt} {\kern 1pt} {\kern 1pt} {\kern 1pt} {\kern 1pt} {\kern 1pt} {\kern 1pt} {\kern 1pt} {\kern 1pt} {\kern 1pt} \left| {{p_i} - {p_j}} \right| \ge 2Q - 1,{\kern 1pt} {\kern 1pt} \forall i,j,i \ne j},
\end{array}{\kern 1pt}
  \end{equation}
where $\left| {{p_i} - {p_j}} \right| \ge 2Q - 1,{\kern 1pt} {\kern 1pt} \forall i,j,i \ne j$ must be met to the establishment of (\ref{eq_lastcompact}). Here, we design the pilot location with given pilot entries to minimize the coherence $\mu(\mathbf{\Phi})$. Instead of exhaustive search, we found that the discrete stochastic optimization (DSO) technique \cite{DSO} can optimize an objective function which can't be evaluated analytically over a collection of feasible parameters. A DSO based value pilot pattern design algorithm is proposed in \cite{pengchengTcom} to seek the optimal pilot pattern within a single symbol. We can extend the algorithm to the joint multi-symbols pilot pattern by simply increasing the dimension, and obtain the optimal value pilot allocation $\mathcal P_{val}$. For simplicity, we give no more redundant illustration of the DSO algorithm.

We will solve the second problem in next section. Here, we would like to remind the readers that the sparse vectors $\{\mathbf s_q\}_{q=0}^{Q-1}$ in (\ref{eq_lastcompact}) could also be recovered using the conventional DCS theory or CS theory.

To this end, our goal is to identify the sparse vectors $\{\mathbf s_q\}_{q=0}^{Q-1}$. For a channel with sparsity $K$, $QG~(Q\ll N, JK<G\ll JL)$ pilot subcarriers are sufficient to estimate the channel within $J$ consecutive OFDM symbols.

\section{Proposed Channel Estimation Algorithm}
In this section, a novel BSOMP algorithm is first proposed to compute the channel parameters. Then, we use novel smoothing treatments based on piecewise linear approximation to reduce the modeling error.
\subsection{The Proposed BSOMP Algorithm}
  Let us define $\mathbf S\buildrel \Delta \over =(\mathbf s_{0},\ldots,\mathbf s_{Q-1})\in \mathbb C^{JL\times Q}$. Considering the joint-block-sparse structure in $\{\mathbf s_q\}_{q=0}^{Q-1}$, we are able to obtain the enhanced distributed compressive channel estimate by solving a $L_0$-norm optimization problem, presented as follows
  \begin{equation}
    \mathbf{ \hat{S}}=arg~min~\|\mathbf u\|_0, ~~s.t. ~\|\mathbf Y -\mathbf \Phi \mathbf S\|_2\leq \varepsilon,
  \end{equation}
  where the vector $\mathbf u=(\|\mathbf S^0\|_2,\ldots,\|\mathbf S^L\|_2)^T\in \mathbb R^{L\times 1}$ and $\mathbf{S}^l=(\mathbf s_0^l,\ldots,\mathbf s_Q^l)\in \mathbb C^{J\times Q}$ is the subblock of the matrix $\mathbf S$. For the reason that we have additional block structural constraint on $\mathbf S$, the SOMP algorithm for conventional DCS needs to be adapted to obtain a more accurate solution, leading to the following block-based SOMP (BSOMP) algorithm.

\begin{algorithm}
\caption{Block-based Simultaneous Orthogonal Matching Pursuit for Channel estimation}
\label{algo:SOMP}
\begin{algorithmic}[1]
\REQUIRE ~~\\
{Received signals: $\mathbf{Y}=([\mathbf{Y}]_{\mathcal{P}_{0}},\ldots,[\mathbf{Y}]_{\mathcal{P}_{Q-1}})$;\\
 Measurement matrix:  $\mathbf{\Phi}  = ({\mathbf{\Phi} _0}, \ldots, {\mathbf{\Phi} _{L - 1}})$;\\
 Sparsity: $K$.}
\ENSURE ~~\\
{$\mathbf{S}=(\mathbf s_{0},\ldots,\mathbf s_{Q-1})$}.
\STATE Initialize the iteration index $i=0$, the sparse vector $\mathbf{S}^{0}=\mathbf 0_{JL \times Q}$, the residual $\mathbf{r}^{0}=\mathbf{Y}-\mathbf{\Phi} \mathbf{S}^{0}=\mathbf{Y}$, the support vector $\mathbf{\Omega}  = {[\mathbf{\Omega} _0^T, \ldots ,\mathbf{\Omega} _{L - 1}^T]^T} = {[\mathbf{0}_J^T, \ldots ,\mathbf{0}_J^T]^T}$ with length $JL$.
\STATE Calculate the residual errors for all $l\in \{0,\ldots,L-1\}$ as $\epsilon_{l}^{i}=\|\mathbf{r}^{i}-\mathbf{\Phi} _{l}{(\mathbf{\Phi} _l^H{\mathbf{\Phi} _l})^{ - 1}}\mathbf{\Phi} _l^H{\mathbf{r}^i}\|_{2}^{^{2}}$.
\STATE Among $\{\epsilon_{l}^{i}\}_{l=0}^{L-1}$ calculated above, find the index $m$ with the minimal residual error $\epsilon_{m}^{i}$. Then update the support vector $\mathbf{\Omega}$ by $\mathbf{\Omega}_{m}=\mathbf{1}_{J}$.
\STATE Update $\mathbf{\Phi} _{\mathbf{\Omega}}$ by extracting the columns of $\mathbf{\Phi}$ according to the updated support vector $\mathbf{\Omega}$. And update the residual as $\mathbf{r}^{i}=\mathbf{Y}-\mathbf{\Phi} _{\mathbf{\Omega}}{(\mathbf{\Phi} _\mathbf{\Omega}^H{\mathbf{\Phi} _\mathbf{\Omega}})^{ - 1}}\mathbf{\Phi} _\mathbf{\Omega}^H{\mathbf{Y}}$.
\STATE $i\leftarrow i+1$.
\STATE Repeat Steps 2 to 5 until $i>K$.
\STATE Based on the optimal least square (LS) estimate, we obtain $\mathbf{S}_{\mathbf{\Omega}}={(\mathbf{\Phi} _\mathbf{\Omega}^H{\mathbf{\Phi} _\mathbf{\Omega}})^{ - 1}}\mathbf{\Phi} _\mathbf{\Omega}^H{\mathbf{Y}}$. Then \textbf{S} is calculated as $\mathbf{S}(\mathbf{\Omega})=\mathbf{S}_{\mathbf{\Omega}}$, while the coefficient vectors out of the support are denoted as $\mathbf{S}(\tilde{\mathbf{\Omega}})=0$.
\end{algorithmic}
\end{algorithm}

In each iteration of \textbf{Algorithm 1}, we first calculate the residual errors for all $l\in \{0,\ldots,L-1\}$ by Step 2. Then, we search the optimal index to make the residual error minimal, and add the corresponding index block to the current support set by Step 3. Note that we update $J$ entries of the support vector simultaneously. In Step 4 we update the measure matrix $\mathbf \Phi$ at the resolution of submatrix with $J$ column vectors, which is different from the SOMP algorithm that only updates one column in each iteration.

  Note that $Q$ sparse coefficient vectors $\{\mathbf s_q\}_{q=0}^{Q-1}$ could also be recovered based on conventional DCS theory with SOMP algorithm, or recovered individually based on the CS theory with OMP algorithm. However, SDCS-BSOMP based scheme can significantly improve the recovery accuracy compared with CS-OMP and DCS-SOMP based schemes, which is due to the fact that the explicit use of joint-block-sparsity in SDCS-BSOMP based scheme improves the success rate in searching the location of nonzero values.

  After recovering coefficient vectors $\{\mathbf s_q\}_{q=0}^{Q-1}$ by \textbf{Algorithm 1}, we can calculate the CE-BEM coefficients $\{\mathbf c_q^{(j)}\}_{q=0}^{Q-1}~(j\in [0,J-1])$ based on (\ref{s1}), (\ref{s2}), and further obtain $\{\mathbf h_l^{(j)}\}_{l=0}^{L-1}$ according to (\ref{ce-bem}). In next subsection, we will carry out smoothing treatment to the already estimated channel tap $\mathbf h_l^{(j)}$ to reduce the CE-BEM modeling error.

\subsection{Smoothing Treatment}

 The CE-BEM could introduce large modeling error, making it difficult to approximate DS channels accurately. In order to reduce the modeling error, we present two smoothing treatment methods to the already estimated channel tap $h_{n,l}$. The first method is carried out within a single OFDM symbol and is related to single-symbol channel estimation scheme, while the second one is performed within multiple OFDM symbols and is related to joint multi-symbols channel estimation scheme. Both of our methods are based on piecewise linear approximation model, which has been proved in \cite{Linear2005} to be a good estimate of DS channel even for normalized Doppler of up to 0.2.

\subsubsection{Piecewise Linear Smoothing within a Single OFDM Symbol}
\begin{figure}[!t]
    \centering
    \includegraphics[width=7cm]{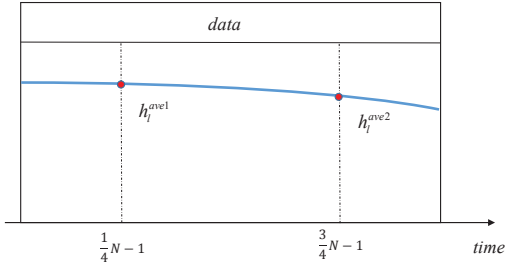}
    \caption{Piecewise linear model within a single OFDM symbol}
    \label{fig.3}
    \end{figure}

  We can approximate the CIR of each subchannel by a linear model during one OFDM symbol. For the sake of simplicity, we drop the index $j$. Let us define $ h_l^{ave1} \buildrel \Delta \over =  E\left( { h_ {0,l} , \ldots , h_ {N/2 - 1,l} } \right)$, $ h_l^{ave2} \buildrel \Delta \over = E\left( { h_ {N/2,l} , \ldots , h_ {N - 1,l} } \right)$, where $h_{n,l}$ is the already estimated channel tap. Considering the linear model proposed in \cite{Linear2005}, we can approximate $h_l\left( {\frac{N}{4} - 1} \right)$ and $h_l\left( {\frac{3N}{4} - 1} \right)$ with the estimate of $ h_l^{ave1}$ and $ h_l^{ave2}$, respectively (see Fig.~\ref{fig.3}). Thus the discrete time gap between $ h_l^{ave1}$ and $ h_l^{ave2}$ is $N/2$ and the slope of the $l$-th path in the current OFDM symbol can be calculated as follows:
  \begin{equation}
    { \alpha _l} = \frac{{ h_l^{ave2} -  h_l^{ave1}}}{{N/2}},{\kern 1pt} {\kern 1pt} {\kern 1pt} {\kern 1pt} {\kern 1pt} {\kern 1pt} {\kern 1pt} {\kern 1pt} {\kern 1pt} {\kern 1pt}  l\in [0,L-1].
  \end{equation}

  Learned the knowledge above, the CIR of the $l$-th path at time $n$ via linear smoothing treatment can easily be derived as
    \begin{equation}\label{eq_line1}
    h_{n,l} = \left( {n + 1 - \frac{N}{4}} \right){ \alpha _l} +  h_l^{ave1},{\kern 1pt} {\kern 1pt} {\kern 1pt} {\kern 1pt} {\kern 1pt} {\kern 1pt} {\kern 1pt} {\kern 1pt} {\kern 1pt} {\kern 1pt} n\in [0,N-1].
    \end{equation}

\subsubsection{Piecewise Linear Smoothing within Multiple OFDM Symbols}
    We define $\{ h_l^{\left( j \right)ave}\}_{l=0}^{L-1}$ to denote the time average of the already estimated CIR during the $j$-th OFDM symbol, represented as
    \begin{equation}
 h_l^{\left( j \right)ave} = \frac{1}{N}\sum\nolimits_{n = j\left( {N + {L_{CP}}} \right) + {L_{CP}}}^{\left( {j + 1} \right)\left( {N + {L_{CP}}} \right) - 1} { h_{n,l} },~~l\in [0,L-1].
    \end{equation}

\begin{figure}[!t]
    \centering
    \includegraphics[width=9cm]{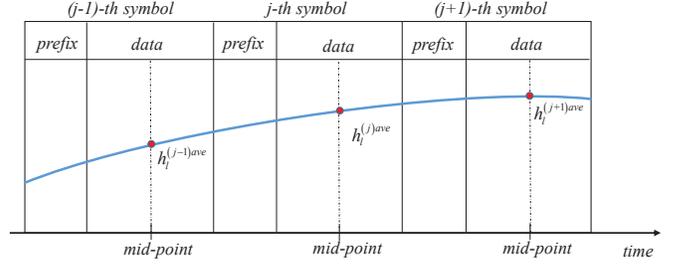}
    \caption{Piecewise linear model within multiple OFDM symbols}
    \label{fig.4}
    \end{figure}

    A significant finding in \cite{Linear2005} is that when $n=(\frac{N}{2}-1)$, $| h_l^{\left( j \right)ave}- h_l^{\left( j \right)}(n)|$ meets its minimum. Consequently, for the $l$-th path, we can approximate $h_l^{\left( j \right)}\left( {\frac{N}{2} - 1} \right)$ with the estimate of $h_l^{\left( j \right)ave}$, which is shown in Fig. \ref{fig.4}. Obviously, we learn that the discrete time gap between $ h_l^{\left( j \right)ave}$ and $ h_l^{\left( j-1 \right)ave}$ is $(N+L_{CP})$, so the estimate of the slope between the $(j-1)$-th and the $j$-th OFDM symbol can be obtained as follows
    \begin{equation}
     \alpha _l^{\left( {j - 1} \right)} = \frac{{ h_l^{\left( j \right)ave} -  h_l^{\left( {j - 1} \right)ave}}}{{N + {L_{CP}}}},~~l\in [0,L-1].
    \end{equation}
    Similarly, the slope between the $j$-th and the $(j+1)$-th OFDM symbol can be obtained as follows
    \begin{equation}
     \alpha _l^{\left( j \right)} = \frac{{ h_l^{\left( j+1 \right)ave} -  h_l^{\left( j \right)ave}}}{{N + {L_{CP}}}},~~l\in [0,L-1].
    \end{equation}

    We can utilize both $ \alpha _l^{\left( {j - 1} \right)}$ and $ \alpha _l^{\left( {j} \right)}$ to calculate the CIR of the $l$-th path, denoted as $\mathbf h_l^{(j)r1}$ and $\mathbf h_l^{(j) r2}$, respectively.
    \begin{equation}\label{eq_line2}
\left\{ {\begin{array}{*{20}{c}}
h_l^{(j)r1}(n)= \left( {n + {L_{CP}} + 1 + \frac{N}{2}} \right) \alpha _l^{\left( {j - 1} \right)} +  h_l^{\left( {j - 1} \right)ave}\\
h_l^{(j)r2}(n)= \left( {n + 1 - \frac{N}{2}} \right) \alpha _l^{\left( j \right)} +  h_l^{\left( j \right)ave}~~~~~~~~~~~~~~
\end{array}} \right.,
    \end{equation}
    where $0\leq n\leq N-1$. Then by calculating the average of $\mathbf h_l^{(j)r1}$ and $\mathbf h_l^{(j)r2}$, we obtain more accurate CIR via piecewise linear smoothing treatment, represented as
    \begin{equation}\label{eq_line3}
    \mathbf h_l^{(j)} = \frac{1}{2}\left( \mathbf h_l^{(j)r1} + \mathbf h_l^{(j)r2} \right){\kern 1pt},~l\in [0,L-1].
    \end{equation}

    We will show in the simulation results in Section \uppercase\expandafter{\romannumeral 5} that the smoothing treatment in (\ref{eq_line1}) and (\ref{eq_line3}) can significantly improve the DS channel estimation performance.

 \subsection{Algorithms Summary and Complexity Analysis}
 Now, we make a summary of our proposed SDCS based joint multi-symbols channel estimation scheme. We first utilize the DSO based value pilot pattern design algorithm proposed in \cite{pengchengTcom} to obtain the optimal value pilot allocation $\mathcal P_{val}$. Then we estimate the coefficient vectors $\{\mathbf s_q\}_{q=0}^{Q-1}$ based on \textbf{Algorithm 1} and calculate the CE-BEM coefficients $\{\mathbf c_q^{(j)}\}_{q=0}^{Q-1}~(j\in [0,J-1])$ based on (\ref{s1}), (\ref{s2}). Next, according to (\ref{ce-bem}), we can obtain $\{\mathbf h_l^{(j)}\}_{l=0}^{L-1}~(j\in [0,J-1])$. Finally, we carry out the smoothing treatment to the already estimated channel tap $\mathbf h_l^{(j)}$ based on piecewise linear approximation by (\ref{eq_line1}), (\ref{eq_line2}), (\ref{eq_line3}), and obtain the final results for channel estimation.

   Here, we briefly discuss the computational complexity of our proposed scheme. Obviously, the main computational burden comes from \textbf{Algorithm 1}. In Step 2, owing to the priori information of $\mathbf \Phi_l$, the complexity of calculating $\mathbf{\Phi} _{l}{(\mathbf{\Phi} _l^H{\mathbf{\Phi} _l})^{ - 1}}\mathbf{\Phi} _l^H$ could be omitted. Then for each iteration, Step 2 can be implemented with the complexity in the order of $\mathcal O\left(G^2Q\right)$. In step 4, we obtain the least square (LS) solution and perform the residual update with the complexity of $\mathcal O\left(GJ^2K^2+J^3K^3+G^2Q\right)$ and $\mathcal O\left(GJKQ\right)$, respectively. Thus, the total complexity of the BSOMP algorithm with $K$ iterations is $\mathcal O\left(GJ^2K^3+J^3K^4+G^2KQ+GJK^2Q\right)$ for $J$ consecutive OFDM symbols. In a practical application, $J$, $K$ and $Q$ are constant parameters and much smaller than $G$. Consequently, we obtain the approximate complexity of \textbf{Algorithm 1} in the order of $\mathcal O\left(G^2\right)$.

\section{Simulation Results and Discussion}
    In this section, simulation studies are performed to show the advantage of our proposed channel estimation scheme. First, based on the joint multi-symbol channel estimation model, we compare the performance of the proposed SDCS scheme with conventional DCS and CS schemes. Then, we present comparisons between our proposed joint multi-symbols channel estimation scheme and conventional single-symbol channel estimation scheme presented in \cite{pengchengTcom}.

    For the simulations, we generate DS channels conforming Jakes' Doppler profile. The parameters of OFDM symbols are based on LTE standard \cite{LTE}, listed in Table \uppercase\expandafter{\romannumeral 1}.

\begin{table}[!ht]
\small
\renewcommand\arraystretch{1}
  \centering
  \caption{PARAMETERS OF THE SIMULATION}
  \begin{tabular}{c|c}
  \hline
  Parameters & Values \\
  \hline
  \hline
  Number of subcarriers & $N = 512$ \\
  Length of CP & $L_{CP}=64$ \\
  Length of CIR & $L=64$ \\
  Nonzero taps & $K=6$\\
  Subcarrier spacing & $\Delta f=15~KHz$ \\
  Bandwidth& $B=7.68~MHz$\\
  CE-BEM order & $Q=3$ \\
  Carrier frequency & $f_c=3~GHz$ \\
  Modulation & QPSK \\
  \hline
\end{tabular}
\end{table}

  The sparse multiple channel $h_{n,l}$ has $K=6$ nonzero channel taps, which are randomly distributed among $L=64$ taps. The channel gain of each path is assumed to obey complex Gaussian distributed according to $\mathcal{CN}(0,\frac{1}{K})$. The variation of the channel is characterized by the normalized Doppler shift (NDS), calculated as $v_{Dmax}=\frac{f_cv}{c\bigtriangleup f}$. To be able to approximate the DS channel by a CE-BEM, we use the standard rule of thumb $Q\geq 2v_{Dmax}+1$ to satisfy the Nyquist criterion. In all simulation cases, we have $v_{Dmax}\leq 1$ such that $Q=3$ could be adequate \cite{zijian2007}. Further, we set the average power of the pilots to be equal to the average power of the data symbols.

   In order to satisfy (13) which guarantees the $J$ consecutive OFDM symbols sharing the same path delays, we set the number of multiple OFDM symbols that are jointly estimated to be $J=3$. The average number of pilot subcarriers within a OFDM symbol in our proposed joint multi-symbol estimation scheme is $P=(2Q-1)G/J=5\times20=100$ with $G=60$. However, in the conventional single-symbol estimation scheme \cite{pengchengTcom}, the number of pilot subcarriers needed is fixed to $P=120$.

 To qualify the channel estimation performance, we calculate the normalized mean square error (NMSE) of different estimators, which is expressed as
  \begin{equation}
    {\rm{NMSE}_{\bar \mathbf h}}\left( {dB} \right) = 10{\log _{10}}\left( {\frac{{E\left( {\left\| {\bar \mathbf h - \bar \mathbf h_{estimated}} \right\|_2^2} \right)}}{{E\left( {\left\| {\bar \mathbf h} \right\|_2^2} \right)}}} \right)£¬
  \end{equation}
  where $\bar \mathbf h_{estimated}$ denotes the estimated CIR. Note that in the above criterion, the true channel $\bar \mathbf h$ is used.

\subsection{NMSE Comparison between SDCS and DCS/CS}
\begin{figure}[!t]
\centering
\includegraphics[width=9cm,height=7cm]{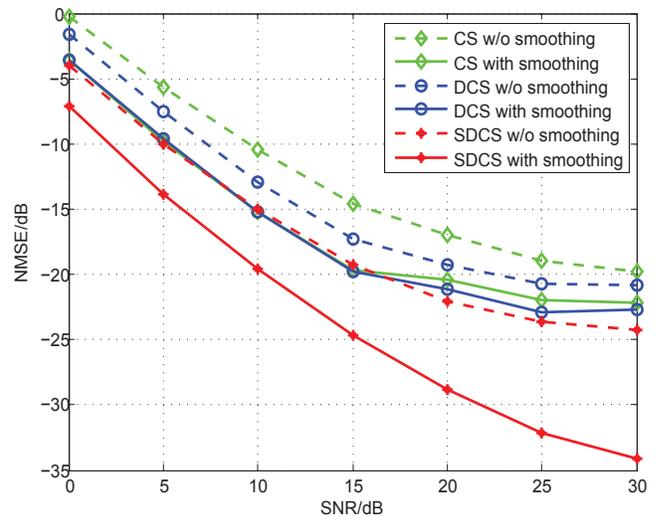}
\caption{Comparison of the NMSE performance between the proposed SDCS scheme and the conventional DCS/CS scheme with $350~km/h$}
\label{fig.5}
\end{figure}

\begin{figure}[!t]
\centering
\includegraphics[width=9cm,height=7cm]{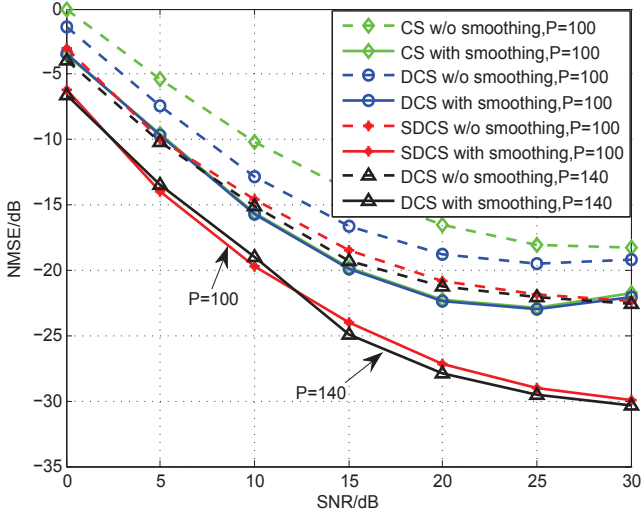}
\caption{Comparison of the NMSE performance between the proposed SDCS scheme and the conventional DCS/CS scheme with $500~km/h$}
\label{fig.6}
\end{figure}

In Figs. \ref{fig.5}-\ref{fig.6}, we show the NMSE comparison of different estimators based on the joint multi-symbol channel estimation model. Meanwhile, the channel estimation accuracy improvement due to the piecewise linear smoothing treatment within multiple OFDM symbols is also demonstrated.

In Fig.~\ref{fig.5}, We have the speed $v=350~km/h$ ($v_{Dmax}=0.065$). Considering the curves without smoothing, it is clearly shown that the SDCS scheme is superior to the DCS and CS schemes. For example, at $\rm {NMSE}=-20$ dB, the SDCS scheme achieves a signal to noise ratio (SNR) gain of about 6 dB compared with DCS scheme and 13 dB compared with CS scheme.

Similar to our expectation, the smoothing treatment within multiple OFDM symbols can reduce the CE-BEM modeling error and improve the performance of channel estimation accuracy. It can be observed in Fig.~\ref{fig.5} that the NMSE performance is significantly improved by smoothing. For example, at $\rm {NMSE}=-20$ dB, the SDCS scheme achieves an SNR gain of around 6 dB from the smoothing treatment.

 In order to further illustrate the better performance of our proposed SDCS scheme for higher vehicle speed, we carry out the similar comparison in Fig.~\ref{fig.6} with the speed of $v=500~km/h~(v_{Dmax}=0.093)$. A similar superiority of our proposed SDCS scheme can be observed. And we can also observe the good performance of the smoothing treatment. However, compared with all curves in Fig.~\ref{fig.5}, the performance of the corresponding curves in Fig.~\ref{fig.6} are degraded, which is mainly due to the CE-BEM modeling error getting larger when Doppler shift increases. In Fig. 6, we also depict curves of the DCS scheme with the average number of pilots $P=140$. It is shown that with the similar NMSE performance between SDCS and DCS, the SDCS scheme incurs an overhead of $\eta=19.5\%$, while the DCS scheme has an overhead of $\eta=27.3\%$. This clearly demonstrates the superiority of our proposed SDCS scheme over the DCS scheme in terms of overhead.

\subsection{NMSE Comparison between Joint Multi-symbol Channel Estimation and Single-symbol Channel Estimation}
In this subsection, we make the NMSE comparison between our proposed joint multi-symbol channel estimation scheme and the conventional single-symbol channel estimation scheme. Here, our proposed joint multi-symbol scheme represents the SDCS-based method combined with piecewise linear smoothing within multiple OFDM symbols, while the single-symbol scheme denotes the channel estimation scheme in \cite{pengchengTcom} combined with piecewise linear smoothing within a single OFDM symbol or combined with a smoothing treatment via discrete prolate spheroidal sequences (DPSSs) \cite{pengchengTcom}. As a reference, we also plot the curves without smoothing treatment.

\begin{figure}[!t]
\centering
\includegraphics[width=9cm,height=7cm]{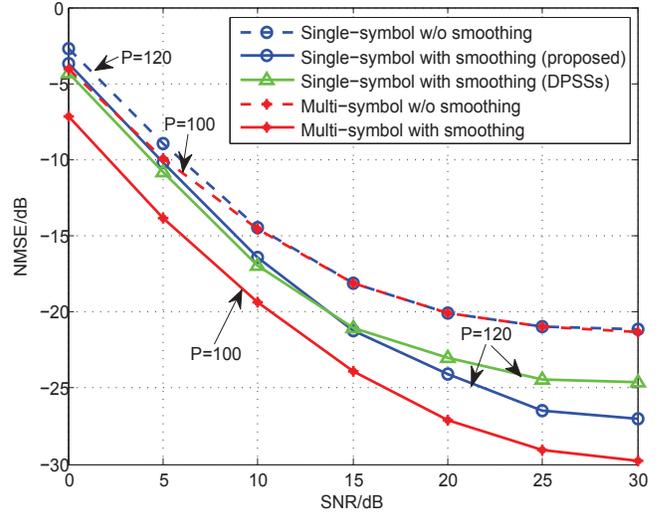}
\caption{Comparison of the NMSE between the joint multi-symbol scheme and the conventional single-symbol scheme with $500~km/h$}
\label{fig.7}
\end{figure}

In Fig. \ref{fig.7}, we carry out the comparison in the condition of $v=500~km/h$. It is clearly shown that the proposed joint multi-symbol scheme significantly outperforms the single-symbol scheme. For example, at $\rm{NMSE}=-20$ dB, the proposed joint multi-symbol scheme achieves an SNR gain of around 3 dB compared with the single-symbol scheme. Furthermore, the single-symbol scheme incurs the number of pilots $P=120$ within an OFDM symbol. In contrast, the proposed joint multi-symbol scheme only has the average number of pilots $P=100$, which represents an improvement in overhead. This apparently demonstrates the superiority of our proposed joint multi-symbol scheme over the single-symbol scheme in terms of both estimation accuracy and spectral efficiency.

We can also observe from Fig. \ref{fig.7} that as to the single-symbol scheme, the proposed smoothing treatment is superior to the method via DPSSs proposed in \cite{pengchengTcom} in the condition of $\rm{SNR}>15$ dB, and both the methods share the similar performance at low SNR. Due to this reason, we only consider the single-symbol scheme combined with the proposed smoothing treatment in the following simulation.


\begin{figure}[!t]
\centering
\includegraphics[width=9cm,height=7cm]{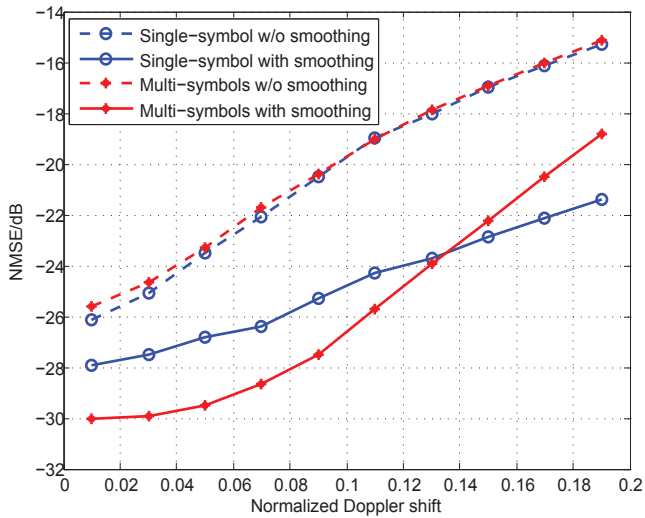}
\caption{NMSE performance versus NDS with $\rm{SNR}=20~dB$}
\label{fig.8}
\end{figure}

To see how Doppler shift influences the performance of channel estimation, Fig. \ref{fig.8} shows the NMSE performance versus the NDS for the case of $\rm{SNR}=20$ dB. The graph shows that with the increase of the NDS, the NMSE curves rise, which is mainly caused by the CE-BEM modeling error getting larger when Doppler shift increases. It can also be observed from Fig. \ref{fig.8} that the proposed joint multi-symbol scheme outperforms the single-symbol scheme when the NDS is less than 0.13 (the speed $v=702~km/h$), while we obtain the opposite conclusion when the NDS is greater than 0.13. This is mainly caused by the fact that the piecewise linear smoothing within multiple symbols fails to approximate the DS channel when the Doppler shift gets big enough. For practical situations, the vehicle speed is usually less than 702 km/h, so our proposed joint multi-symbol scheme is meaningful.

\subsection{BER Performance}
\begin{figure}[!t]
\centering
\includegraphics[width=9cm,height=7cm]{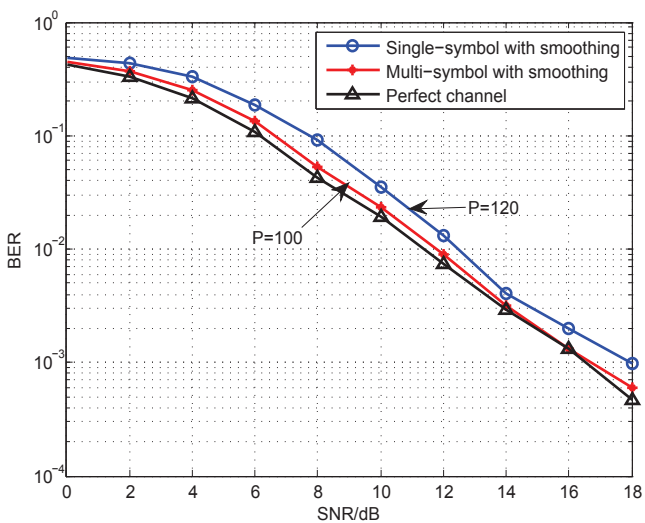}
\caption{Comparison of the coded BER performance between the joint multi-symbol scheme and the conventional single-symbol scheme with $500~km/h$}
\label{fig.9}
\end{figure}

In Fig. \ref{fig.9}, we compare the coded bit error rate (BER) performance of the proposed joint multi-symbol scheme with the single-symbol scheme. As a reference, we also plot the BER performance under the ideal channel, which means that $\mathbf H_F$ in (7) is available at the receiver. Here, the rate-1/2 convolutional code is applied and the zero-forcing (ZF) equalizer is adopted. It is clearly shown that the proposed joint multi-symbol scheme significantly outperforms the single-symbol scheme. For example, at $\rm{BER}=10^{-2}$, the proposed joint multi-symbol scheme achieves an SNR gain of around 1 dB compared with the single-symbol scheme, and is only about 0.2 dB away from the case with perfect channel knowledge.

\section{Conclusion}
  In this paper, we presented a novel SDCS based joint multi-symbol channel estimation scheme over a DS channel. By utilizing the CE-BEM and designing a special sparse pilot pattern within multiple OFDM symbols, we transformed the original sparse DS channel into a joint-block-sparse channel model, and proposed a novel BSOMP algorithm to exploit the jointly sparse block structure of the coefficient vectors. To reduce the modeling error induced by the CE-BEM, two smoothing treatment methods via piecewise linear approximation were proposed. Simulation results demonstrate the proposed SDCS-based scheme achieves higher estimation accuracy than the conventional DCS-based and CS-based scheme when tracking the joint multi-symbol estimation model, and the proposed joint multi-symbol scheme outperforms the conventional single-symbol scheme in terms of both estimation accuracy and spectral efficiency.

\appendix  
For convenience, we temporarily ignore $\mathbf{Z}$ of illustration. From (7), we have
\begin{equation}
    \mathbf{Y}^{(j)} = \sum\limits_{q = 0}^{Q-1} {\mathbf{I}_N^{\langle q-\frac{Q-1}{2} \rangle}\mathbf{\bm{\mathcal{D}}}\left\{ {\mathbf{X}^{(j)}} \right\}} {{{\mathbf{V}_L}\mathbf{c}_q^{(j)}}}.
  \end{equation}
Then (\ref{eq_mul}) can be rewritten as
 \begin{equation}\label{eq_Y}
    \mathbf{Y} = \sum\limits_{q = 0}^{Q-1} {\mathbf{I}_{JN}^{\langle q-\frac{Q-1}{2} \rangle}\mathbf{\bm{\mathcal{D}}}\left\{ {\mathbf{X}} \right\}} {{({\mathbf I_J} \otimes{\mathbf{V}_L}){\left( {\begin{array}{*{20}{c}}
    {\mathbf c_{q}^{(0)}}\\
     \vdots \\
    {\mathbf c_{q}^{(J-1)}}
    \end{array}} \right)}}}.
  \end{equation}

  Let us define the $q$-th $(q\in [0,Q-1])$ pilot subcarriers selector matrix as $\mathbf{R}_q \buildrel \Delta \over = [\mathbf{I}_{JN}]_{{\mathcal P_{q}}}\in \mathbb C^{G\times JN} $. It then follows from (\ref{eq_Y}) that
  \begin{equation}
    [\mathbf{Y}]_{{\mathcal P_{q}}} = \mathbf{R}_q\sum\limits_{q' = 0}^{Q-1} {\mathbf{I}_{JN}^{\langle q'-\frac{Q-1}{2} \rangle}\mathbf{\bm{\mathcal{D}}}\left\{ {\mathbf{X}} \right\}} {{({\mathbf I_J} \otimes{\mathbf{V}_L}){\left( {\begin{array}{*{20}{c}}
    {\mathbf c_{q'}^{(0)}}\\
     \vdots \\
    {\mathbf c_{q'}^{(J-1)}}
    \end{array}} \right)}}}.
    \end{equation}
  Due to ${\mathbf R_q}\mathbf I_{JN}^{\left\langle {q' - \frac{{Q - 1}}{2}} \right\rangle } = {\mathbf R_{q - q' + \frac{{Q - 1}}{2}}}$, then we obtain
\begin{equation}\label{eq_Ypq}
    [\mathbf{Y}]_{{\mathcal P_{q}}}=\sum\limits_{q' = 0}^{Q-1} {\mathbf{R}_{q-q'+\frac{Q-1}{2}}\mathbf{\bm{\mathcal{D}}}\left\{ {\mathbf{X}} \right\}} {{({\mathbf I_J} \otimes{\mathbf{V}_L}){\left( {\begin{array}{*{20}{c}}
    {\mathbf c_{q'}^{(0)}}\\
     \vdots \\
    {\mathbf c_{q'}^{(J-1)}}
    \end{array}} \right)}}}.
  \end{equation}
  Considering the pilot subcarriers
  \begin{equation}
{\left[ \mathbf X \right]_{{\mathcal P_q}}} = \left\{ {\begin{array}{*{20}{c}}
{{\mathbf P_{val}}{\kern 1pt} {\kern 1pt} {\kern 1pt} {\kern 1pt} {\kern 1pt} {\kern 1pt} {\kern 1pt} q = \frac{{Q - 1}}{2}}\\
{\mathbf 0{\kern 1pt} {\kern 1pt} {\kern 1pt} ~~{\kern 1pt} {\kern 1pt} {\kern 1pt} {\kern 1pt} {\kern 1pt} {\kern 1pt} {\kern 1pt} {\kern 1pt} {\kern 1pt} {\kern 1pt} q \ne \frac{{Q - 1}}{2}}
\end{array}} \right.,
  \end{equation}
  we can extract nonzero values from $\mathbf X$ only if $q'=q$. Consequently, (\ref{eq_Ypq}) can be rewritten as
  \begin{equation}
    [\mathbf{Y}]_{{\mathcal P_{q}}} = \mathbf{\bm{\mathcal{D}}}\left\{  {\mathbf P_{val}}\right\} {{[{\mathbf I_J} \otimes{\mathbf{V}_L}]_{{\mathcal P_{val}}}{\left( {\begin{array}{*{20}{c}}
    {\mathbf c_{q}^{(0)}}\\
     \vdots \\
    {\mathbf c_{q}^{(J-1)}}
    \end{array}} \right)}}}
  \end{equation}
  where $0\leq q\leq Q-1$. Until now, we can obtain (\ref{EQ_mulmodel}).


{
\small
\bibliographystyle{ieeetr}
\bibliography{bob}
}

\end{document}